\documentclass[sort&compress,5p]{elsarticle}
\usepackage{epsf}
\usepackage{amsmath}
\usepackage{amsfonts}
\usepackage{amssymb}
\usepackage{mathptmx}

\begin{document}

\title{Anisotropy of electric resistance and upper critical field
in magnetic superconductor Dy$_{0.6}$Y$_{0.4}$Rh$_{3.85}$Ru$_{0.15}$B$_4$  }

\author[ilt,trz,int]{A. V. Terekhov\corref{cor1}}\ead{terekhov1977@yandex.ru}
\author[ilt]{I. V. Zolochevskii}
\author[ilt]{E. V. Khristenko}
\author[ilt]{L. A. Ishchenko}
\author[ilt]{E. V. Bezuglyi}
\author[trz]{A. Zaleski}
\author[int,ver]{E. P. Khlybov}
\author[bai]{S. A. Lachenkov}
\cortext[cor1]{Corresponding author}

\address[ilt]{B.I. Verkin Institute of Low Temperature Physics and
Engineering, National Academy of Sciences of Ukraine, Lenin Av. 47, Kharkiv
61103, Ukraine}
\address[trz]{W. Trzebiatowski Institute of Low Temperatures and Structure
Research, Polish Academy of Sciences, Box 1410, Okolna 2, Wroclaw 50-950,
Poland}
\address[int]{International Laboratory of High Magnetic Fields and Low
Temperatures, Polish Academy of Sciences, ul. Gajowicka 95, Wroclaw 53-421,
Poland }
\address[ver]{L.F. Vereshchagin Institute of High Pressure Physics,
Russian Academy of Sciences, Kaluzhskoe shosse 14, Troitsk 142190, Russia}
\address[bai]{A.A. Baikov Institute of Metallurgy and Metallography,
Russian Academy of Sciences, Leninskii pr. 49, Moscow 119991, Russia}

\begin{abstract}
We have measured temperature dependencies of the electric
resistance $R$ and upper critical magnetic field $H_{c2}$ of a
magnetic superconductor
Dy$_{0.6}$Y$_{0.4}$Rh$_{3.85}$Ru$_{0.15}$B$_4$. The measurements
were made for different angles $\varphi$ of magnetic field
inclination to the direction of measuring current and revealed
strong anisotropy of the behavior of $R(T)$ and the values of
$H_{c2}(T)$. By using the Wert\-hamer-Gelfand-Hohenberg theory, we
determined the Maki parameter $\alpha$ and the parameter of the
spin-orbital interaction. For $\varphi = 0^\circ$ and $90^\circ$
both parameters are close to zero, thus the magnitude of
$H_{c2}(0) \approx 38$ kOe is basically limited by the orbital
effect. At $\varphi = 45^\circ$, a large value of the parameter
$\alpha = 4.2$ indicates dominating role of the spin-paramagnetic
effect in the suppression of $H_{c2}(0)$ down to $8.8$ kOe. We
suggest that such behavior of $R(T)$ and $H_{c2}(T)$ is caused by
internal magnetism of the Dy atoms which may strongly depend on
the magnetic field orientation.
\end{abstract}


\maketitle

\section{Introduction}

Ternary compounds whose structures include a regular sublattice of
magnetic moments are attractive objects for studying the influence
of magnetism on superconductivity. Among these materials, the most
famous are PbMo$_6$S$_8$-type ``Chevrel phases'' and ternary
rare-earth rhodium borides \cite{Maple}. The physical properties
of quadruple compounds Dy$_{1-x}$Y$_x$Rh$_4$B$_4$ having a
body-centered tetragonal LuRu$_4$B$_4$-type crystal structure
deserve special attention due to a great number of interesting
features of these materials. For instance, it was found
\cite{Dmitr1,Dmitr2} that the magnetic ordering of Dy atoms may
occur at the temperature $T_M$ higher than the superconducting
transition temperature $T_c$ and coexist with superconductivity
down to very low temperatures. It was established in \cite{Dmitr2}
that Dy$_{1-x}$Y$_x$Rh$_4$B$_4$ belongs to materials with
intrinsic ferrimagnetism, and the transition temperature $T_M$
strongly depends on the concentration of non-magnetic yttrium: it
falls with increasing Y concentration from 37 K in DyRh$_4$B$_4$
down to 7 K in Dy$_{0.2}$Y$_{0.8}$Rh$_4$B$_4$. On the contrary,
$T_c$ increases with the Y concentration from 4.7 K for
DyRh$_4$B$_4$ to 10.5 K in YRh$_4$B$_4$ \cite{Dmitr1}.
Measurements of the specific heat of
Dy$_{0.8}$Y$_{0.2}$Rh$_4$B$_4$ ($T_M = 30.5$ K, $T_c = 5.9$ K)
indicate emergence of another magnetic phase transition below 2.7
K \cite{Dmitr2}. Recently, anomalies of some physical quantities,
unusual for systems with conventional superconductivity, were
discovered in Dy$_{1-x}$Y$_x$Rh$_4$B$_4$: the paramagnetic
Meissner effect \cite{Dmitr3,Ter2013} and non-monotonic
temperature dependencies of the upper critical magnetic field
$H_{c2}$ and the superconducting gap \cite{Dmitr2,Dmitr4,Ryba}.

Another specific feature of this class of magnetic superconductors
is the change of the type of magnetic interactions in the Dy
subsystem under partial replacement of rhodium by ruthenium. As
shown in \cite{Ham1}, antiferromagnetic ordering in
Dy(Rh$_{1-y}$Ru$_y$)$_4$B$_4$ holds for $y < 0.5$ and changes to a
ferromagnetic one for $y > 0.5$. The superconducting transition
temperature also varies with the Ru concentration \cite{Ham1}.

Thus, the study of physical properties of the borides family
Dy$_{1-x}$Y$_x$(Rh, Ru)$_4$B$_4$ with various content of
dysprosium (responsible for the magnetic interactions) and of
ruthenium and rhodium (responsible for both the magnetic
interactions and superconductivity) is of great interest to
explore the coexistence of superconductivity and magnetism and the
possibility of appearance of unconventional superconductivity. For
this purpose, we studied in this paper the behavior of the
electric resistance in the vicinity of the superconducting
transition and the upper critical field in the compound
Dy$_{0.6}$Y$_{0.4}$Rh$_{3.85}$Ru$_{0.15}$B$_4$ at different
orientations of external magnetic field with respect to the
direction of measuring electric current.

\section{Results and discussion}

The Dy$_{0.6}$Y$_{0.4}$Rh$_{3.85}$Ru$_{0.15}$B$_4$ compound has
been prepared by the argon arc melting of initial components,
followed by annealing within a few days. The resulting
single-phase polycrystalline ingot had a LuRu$_4$B$_4$-type
crystal structure (space group I4/mmm) testified by the X-ray
phase and structural analyses. At this concentration of ruthenium,
it is possible to synthesize samples with such structure at the
normal pressure, in contrast to the quadruple compounds
Dy$_{1-x}$Y$_x$Rh$_4$B$_4$, which gain the required structure only
during the synthesis under a pressure of 8 GPa. The samples were
cut from the ingot in the form of parallelepipeds whose lengths
were about 6 mm and the cross-sectional area was $1 \times 1$ mm.
The measurements of the electrical resistance $R(T)$ of the
samples were performed on a Quantum Design PPMS-9 automatic system
using a standard four-probe circuit with an alternating current
($I = 8$ mA, $f = 97$ Hz) in the temperature range $2-12$ K and
magnetic fields up to 36 kOe produced by a superconducting
solenoid. The sample holder was equipped with a system for
automatic rotation of the substrate with the sample by a stepping
motor of high resolution which allowed the measurement of $R(T)$
for different angles $\varphi$ of inclination of the external
magnetic field $H$ to the direction of the current. The
superconducting transition temperature measured in the middle of
the resistive transition in zero magnetic field was $6.66$ K.

\begin{figure}[tb]
\centerline{\epsfxsize=8.5cm\epsffile{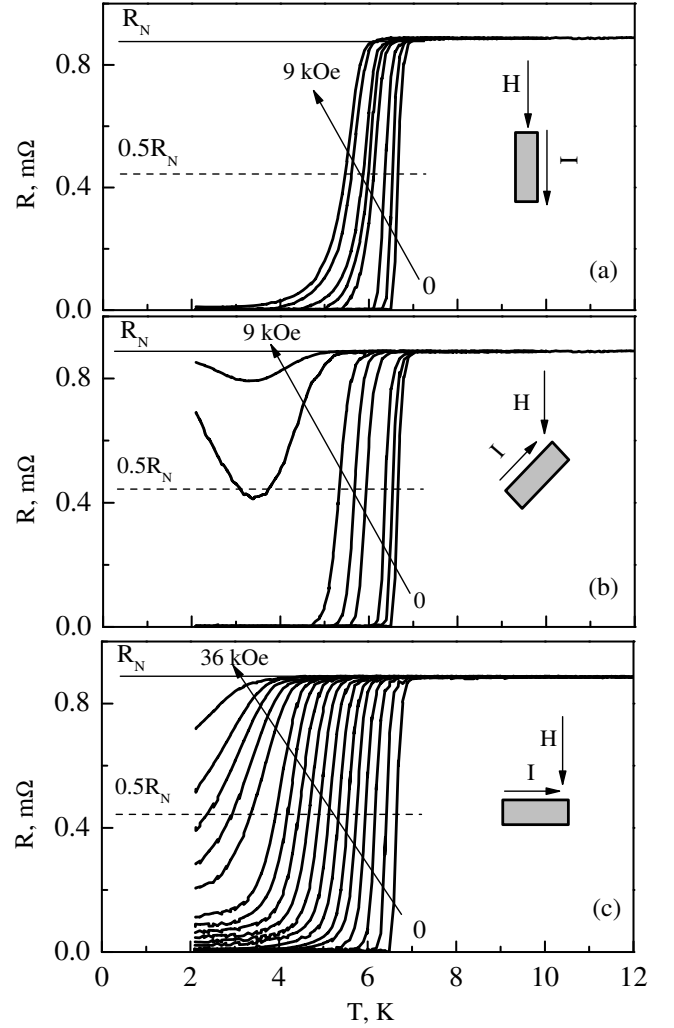}}
\caption{Temperature dependencies of the resistance for three orientations of
the magnetic field relative to the longitudinal sample axis: $\varphi=0^\circ$,
$H \| I$ (a); $\varphi=45^\circ$ (b); $\varphi=90^\circ$, $H \bot I$  (c) in
magnetic fields $0, 1, 2, 4, 5, 6, 8, 9$ kOe for $\varphi = 0^\circ, 45^\circ$
and $0-36$ kOe through 2 kOe for $\varphi = 90^\circ$.}
\label{fig1}
\end{figure}

Fig.\ref{fig1} presents the temperature dependencies of the sample
resistance in different magnetic fields of three orientations: $\varphi =
0^\circ$, $45^\circ$ and $90^\circ$ (panels (a), (b) and (c),
respectively). For the angles $\varphi = 0^\circ$ and $45^\circ$, the
experiments were made in magnetic fields $H=0-9$ kOe and for $\varphi =
90^\circ$ -- at $H=0-36$ kOe. The shape of $R(T)$ in the range of fields
$0-6$ kOe is typical for the superconducting transition: a sharp fall of
the resistance below a certain temperature followed by its disappearance
at lower temperatures. Another type of the behavior of $R(T)$ was observed
at $\varphi = 45^\circ$ in the fields larger than 6 kOe. In this case, the
resistance decreases only down to a certain finite value $R_{min} \approx
0.4R_N$ ($R_N$ is the sample resistance of in the normal state), and then,
as the temperature is lowered further, $R(T)$ rapidly increases. With
further increase of the field, the observed minimum of $R(T)$ shifts to
lower temperatures and reduces in its depth up to $R_{min} \approx
0.9R_N$. Thus it can be argued that the destruction of superconductivity
at the magnetic field orientation $\varphi = 45^\circ$ (Fig.\ref{fig1}b)
begins in the fields much smaller than at the orientations $\varphi =
0^\circ$ and $90^\circ$.

Such strong anisotropy of $R(T)$ is apparently due to the presence of the
texture (preferred orientation of individual crystallites) in a
polycrystalline sample and the coexistence of the magnetic and
superconducting orderings. Along this line of reasonings, the minimum in
$R(T)$ for $\varphi = 45^\circ$ can be attributed to stronger (compared to
other orientations) enhancement of an uncompensated magnetic moment with
increasing magnetic field. This excess magnetism leads to significant
suppression of the superconducting state in the fields $> 8$ kOe at
$\varphi = 45^\circ$, while for other orientations, superconductivity
holds up to several tens of kOe.

\begin{figure}[tb]
\centerline{\epsfxsize=8.5cm\epsffile{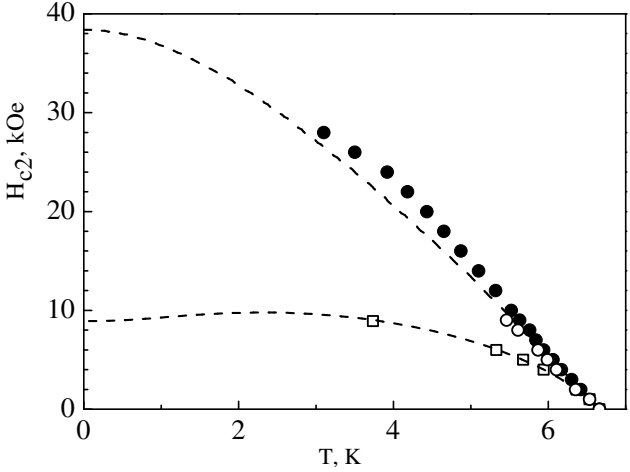}}
\caption{Temperature dependencies of the upper critical magnetic field
$H_{c2}$ for $\varphi = 0^\circ$ ({\Large$\bullet$}), $45^\circ$ ($\square$)
and $90^\circ$ ({\Large$\circ$}). Dashed lines show the results of the WHH
theory with fitting parameters of the spin-paramagnetic and spin-orbital
interaction.}
\label{fig2}
\end{figure}

We note that a minimum of $R(T)$ in magnetic fields has been
earlier observed in other magnetic superconductors such as
NdRh$_4$B$_4$ \cite{Ham} and Dy$_{1.2}$Mo$_6$S$_8$ \cite{Ishi}. It
has been attributed to the induction of the magnetic ordering of
Nd and Dy ions by an external magnetic field at the transition
temperature $T_M < T_c$, which leads to destruction of the
superconducting state (reentrant superconductivity). However, in
our case, the superconductivity and the magnetic order, which
emerges at $T_M > T_c$, coexist below $T_c$, and the minimum in
$R(T)$ can be explained as the result of changes in the existing
magnetic structure caused by the magnetic field of specific
orientation.

Using the data of Fig.\ref{fig1} and accepting for $H_{c2}(T)$ the
values of the external magnetic field and the temperature, at
which $R(H,T) = 0.5R_N$, we plotted the experimental temperature
dependencies of the upper critical field depicted by circles and
squares in Fig.\ref{fig2}. In contrast to the previous Andreev
spectroscopy data \cite{Dmitr2,Dmitr4,Ryba}, we did not found any
non-monotony in the behavior of $H_{c2}(T)$ that possibly reflects
certain ambiguity in the interpretation of the results of the
point-con\-tact measurements in nonhomogeneous samples; a certain
role in this difference may be also played by the admixture of
ruthenium in our samples. Dashed curves in Fig.\ref{fig1} show the
dependencies $H_{c2}(T)$ calculated from the equation of the
Wert\-ha\-mer-Gelfand-Ho\-hen\-berg (WHH) theory \cite {WHH}:
\begin{align} \label{WHH}
\ln\frac{1}{t} &= \left(\frac{1}{2}+\frac{i\lambda_{so}}{4\gamma}\right) \psi
\left(\frac{1}{2}+\frac{\bar{h}+\frac{1}{2}\lambda_{so} + i\gamma}{2t}\right)
\\ \nonumber
&+\left(\frac{1}{2}-\frac{i\lambda_{so}}{4\gamma}\right) \psi
\left(\frac{1}{2}+\frac{\bar{h}+\frac{1}{2}\lambda_{so} - i\gamma}{2t}\right).
\end{align}
Here $\psi(x)$ is the digamma function, $\gamma = \sqrt{(\alpha
\bar{h})^2 - (\lambda_{so}/2)^2}$, and
\begin{align} \label{reduced}
&t = \frac{T}{T_c}, \qquad \bar{h} = -\frac{4}{\pi^2}
\frac{H_{c2}}{(dH_{c2}/dt)_{t=1}}
\end{align}
are the reduced temperature and critical magnetic field,
respectively. In our calculations, we use the fitting values of
the Maki parameter $\alpha$ which describes relative contribution
of the spin-paramagnetic effect and the parameter $\lambda_{so}$
of the spin-orbit scattering. The best fit for $\varphi = 0^\circ$
and $90^\circ$ gives $\alpha = \lambda_{so}=0$, i.e., only the
orbital effect is responsible for the suppression of
superconductivity, whereas at $\varphi = 45^\circ$ we obtain a
rather large value of $\alpha=4.2$ ($\lambda_{so}=0$) which
indicates essential contribution of the spin-paramagnetic effect.

Figure 2 shows that for $\varphi=0^\circ$, the experimental values of
$H_{c2}(T)$ at the temperatures below $0.8 T_c$ slightly exceed the
maximum possible calculated ones. This could be explained either by the
effect of anisotropy (may lead to increase in $H_{c2}$ by $20-30\%$
\cite{Hoh}) or by the presence of strong coupling in the superconducting
condensate (can enhance $H_{c2}$ by $1.3$ times \cite{Rai}), which are
beyond the frameworks of the WHH theory. The possibility of the strong
coupling in this material follows from the results of the Andreev
spectroscopy of the superconducting gap $\Delta$ \cite{Ryba}, according to
which the ratio $2\Delta/kT_c$ can reach 4 or even higher values, larger
than the value 3.52 for conventional superconductors with the weak
coupling.

The orbital critical field at $T=0$ can be calculated by using the
formula of the WHH theory for the dirty limit \cite{WHH},
\begin{align} \label{Horb}
&H^{\textit{orb}}_{c2}(0) = -0.69 T_c \,(dH_{c2}/dT)_{T=T_c},
\end{align}
while the upper critical field can be estimated as \cite{Maki}
\begin{align} \label{Hc2}
&H_{c2}(0) = \frac{H^{\textit{orb}}_{c2}(0)}{\sqrt{1+\alpha^2}}.
\end{align}
As is obvious from Fig.\ref{fig2}, initial slopes of $H_{c2}(T)$
near $T_c$ are approximately equal for all orientations of the
magnetic field. According to \eqref{Horb}, this results in a
universal (angle-independent) value of $H^{\textit{orb}}_{c2}(0)
\approx 38$ kOe. Since at $\varphi = 0^\circ$ and $90^\circ$ the
fitting value of the Maki parameter is close to zero, we conclude
that the orbital field at these orientations fully determines the
magnitude $H_{c2}(0) = H^{\textit{orb}}_{c2}(0)$ of the upper
critical field at zero temperature. The estimate of a small
paramagnetic contribution can be obtained from the relation $\mu_B
H^{\textit{p}}_{c2}(0) = 1.84 kT_c$ for the Chandrasekhar-Clogston
limit \cite{Chandra,Clog} ($\mu_B$ is the Bohr's magneton) which
gives the critical paramagnetic field
$H^{\textit{p}}_{c2}(0)=122.5$ kOe. Then, using the equation
\cite{Maki}
\begin{align} \label{alpha}
&\alpha = \sqrt{2} \frac{H^{\textit{orb}}_{c2}(0)}{H^{\textit{p}}_{c2}(0)}
\end{align}
we found a rather small value $\alpha \approx 0.4$ for the Maki
parameter which explains unobservability of the spin effects at
the experimental dependencies $H_{c2}(T)$ at $\varphi = 0^\circ$
and $90^\circ$.

As noted above, at $\varphi = 45^\circ$, the large value of the
Maki parameter $\alpha = 4.2$ implies that the spin-paramagnetic
effect plays the main role in the suppression of
superconductivity. In this case, equation \eqref{Hc2} gives
$H_{c2}(0) \approx 8.8$ kOe, i.e., by 4.3 times smaller than its
value for the orientations $\varphi = 0^\circ$ and $90^\circ$.
Correspondingly, the magnitude of the critical paramagnetic field
$H^{\textit{p}}_{c2}(0)=12.8$ kOe found from \eqref{alpha} appears
to be much smaller than at $\varphi = 0^\circ$ and $90^\circ$.
These results give additional arguments in the benefit of our
assumption about rearrangement of the magnetic structure and
formation of an excess magnetic moment induced by the external
magnetic field. This considerably enhances the effective magnetic
field acting on the electron spins which causes destruction of the
Cooper pairs.

\begin{figure}[tb]
\centerline{\epsfxsize=8.5cm\epsffile{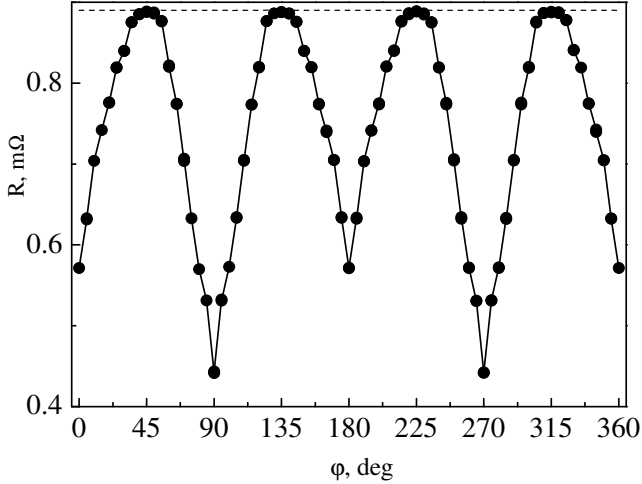}}
\caption{Angle dependencies of the electric resistance within the range
$\varphi = 0^\circ - 360^\circ$ at the temperature $5.75$ K corresponding
to the middle of the superconducting transition in the field of 8 kOe.
Dashed line depicts the angle-independent resistance at $T = 9$ K, $H=8$
kOe (normal state).}
\label{fig3}
\end{figure}

In order to obtain additional information about the effect of the
magnetic field inclination on the suppression of
superconductivity, we measured the angle dependencies of the
sample resistance within the range $\varphi = 0^\circ - 360^\circ$
(see Fig.\ref{fig3}) at the temperature $5.75$ K which corresponds
to the middle of the superconducting transition at $\varphi=0$ in
the field of 8 kOe. Figure \ref{fig3} shows that with an increase
in the angle, the resistance first grows to a maximum value $R_N$
at $\varphi = 45^\circ$, then begins to drop with a minimum at
$\varphi = 90^\circ$. All highs in $R(\varphi)$ repeat themselves
through $90^\circ$, and lows -- through $180^\circ$ (the magnitude
of minimum ohmic losses for $\varphi = 0^\circ$ and $180^\circ$ is
smaller than that at $\varphi = 90^\circ$ and $270^\circ$). Thus,
the superconductivity is most strongly suppressed in the magnetic
fields directed at the angles $45^\circ$ plus multiple of
$90^\circ$ relative to the longitudinal sample axis. The fields
inclined at the angles by multiple of $90^\circ$ have the weakest
impact on the superconducting state. The dashed line in
Fig.\ref{fig3} indicates the experimental data obtained at $T=9$ K
in the field of 8 kOe and demonstrates independence of the sample
resistance of the field direction in the normal state.

\section{Summary}

We have measured the resistance $R$ and the upper critical
magnetic field $H_{c2}$ of the magnetic superconductor Dy$_{
0.6}$Y$_{0.4}$Rh$_{3.85}$Ru$_{0.15}$B$_4$ at different angles
$\varphi$ of inclination of the magnetic field relative to the
longitudinal sample axis. The temperature dependencies $R(T)$ and
$H_{c2}(T)$ are strongly anisotropic in the superconducting state,
whereas the rotation of the magnetic field in the normal state has
no effect on its resistive properties. Suppression of
superconductivity is most pronounced at $\varphi=45^\circ$ plus
multiple of $90^\circ$ ($H_{c2}(0) \approx 8.8$ kOe), while at the
angles $\varphi=0^\circ$ and $90^\circ$ the effect of the magnetic
field on the superconducting state is weakest, and the calculated
magnitude of $H_{c2}(0)$ reaches 38 kOe. A minimum in $R(T)$ in
large enough fields was observed at $\varphi=45^\circ$ which
resembles reentrance effects in some magnetic superconductors near
the point of transition to the magnetically ordered state.
However, in our case, this minimum most likely occurs due to
restructuring of already {existed} magnetic ordering.

Analysis of the behavior of $H_{c2}(T)$ within the framework of
the WHH theory shows that for $\varphi = 0^\circ$ and $90^\circ$
the Maki parameter $\alpha$ and the parameter $\lambda_{so}$ of
the spin-orbit scattering are close to zero, i.e., only the
orbital effect is responsible for the suppression of
superconductivity. This is confirmed by the estimate of $\alpha$
obtained from the calculated paramagnetic limit. On the contrary,
at $\varphi = 45^\circ$, the Maki parameter was found to be large
($\alpha=4.2$, $\lambda_{so}=0$) which manifests the dominating
role of the spin-paramagnetic depairing mechanism. We suggest that
the above mentioned features of the behavior of superconducting
parameters may be associated with the growth of a spontaneous
magnetic moment of the dysprosium subsystem induced by an external
magnetic field of specified orientation. At the same time, one can
not exclude the existence of an unconventional pairing mechanism,
such as a triplet pairing, in this material.


\end{document}